\documentclass[11pt,preprint]{aastex}
\usepackage{epsfig,lscape}
\slugcomment{KSUPT-11/1 \hspace{0.5truecm} May 2011}

\usepackage{natbib}

\newcommand{\lap}{\lower.5ex\hbox{$\; \buildrel < \over \sim \;$}}
\newcommand{\gap}{\lower.5ex\hbox{$\; \buildrel > \over \sim \;$}}

\newcommand{\kmsmpc}{{\rm \, km\, s}^{-1}{\rm Mpc}^{-1}}

\tighten
 
\begin{document}
\title{Median statistics and the Hubble constant}
\author{Gang Chen\altaffilmark{1} and Bharat Ratra\altaffilmark{2}}

\altaffiltext{1}{Tianjin Astrophysics Center, Tianjin Normal University, Tianjin 300387, China, chengump@gmail.com}
\altaffiltext{2}{Department of Physics, Kansas State University, 116 Cardwell
                 Hall, Manhattan, KS 66506, USA, ratra@phys.ksu.edu}

\begin{abstract}
Following Gott et al.\ (2001), we use Huchra's final compilation of 553 
measurements of the Hubble constant ($H_0$) to determine median statistical 
constraints on $H_0$. We find $H_0=68 \pm 5.5$ (or $\pm 1$) $\kmsmpc$, where 
the errors are the 95\% statistical and systematic (or statistical) errors. 
With about two-third more measurements, these results are close to what 
Gott et al.\ found a decade ago, with smaller statistical errors and 
similar systematic errors. 
\end{abstract}


\section{Introduction}

The long and involved history of increasingly more accurate and precise 
measurements of the Hubble constant has resulted in an extensive list 
of more than 550 $H_0$ values recorded by Huchra.\footnote{
See cfa-www.harvard.edu/$\sim$huchra/.} Rowan-Robinson (2009) notes that
most recent (central) estimates of $H_0$ lie in the range of 62 to 72
$\kmsmpc$, although individual estimates can differ amongst themselves 
by 2 or 3 standard deviations (for recent reviews see Jackson 2007;
Tammann et al.\ 2008; Freedman \& Madore 2010). While this is unfortunate,
and perhaps not unexpected, Huchra's extensive compilation may be used
to derive a more precise meta-estimate of $H_0$ that is more robust
than any individual estimate.

We follow Gott et al.\ (2001, hereafter G01) and use median statistics 
to determine what Huchra's $H_0$ central values alone (i.e., ignoring 
the quoted errors) tell us about the true value and uncertainty of 
the Hubble constant.\footnote{The uncertainty in $H_0$ affects the 
uncertainty in other cosmological parameters determined from some 
cosmological tests, see, e.g., Wilson et al.\ (2006), Wan et al.\ (2007),
Samushia et al.\ (2007, 2010), Zhang et al.\ (2007), Sen \& Scherrer 
(2008), and Dantas et al.\ (2011).}  
With about two-third more data than G01 (553 measurements versus 331), 
we confirm and strengthen the results of G01.\footnote{For an
analysis of an intermediate version of Huchra's list with 461 measurements,
see the Appendix of Chen et al.\ (2003). Chen et al.\ (2003) did not estimate
systematic errors bars; instead they used the earlier systematic error 
estimate of G01. In this paper we estimate systematic error bars for 
the new list of $H_0$ measurements.}
We also examine how the estimated value of $H_0$ changes as we consider 
different subsamples of the complete list, and argue that the estimate 
from the complete list is a robust estimate of the Hubble constant.

Our paper is organized as follows. We first review some basic median 
statistics concepts from G01 in the following section. In Sec.\ 3 these
are applied in an analysis of Huchra's $H_0$ list, where we also 
discuss some consistency tests and give constraints on the 
Hubble constant. We conclude in Sec.\ 4.

\section{Median statistics and errors}

Compared to a $\chi^2$ analysis, a median statistics analysis requires 
fewer hypotheses and is much less sensitive to being biased by outliers.
See G01 for a comprehensive introduction to median statistics and its 
applications.\footnote{For further applications and more recent 
developments see Podariu et al.\ (2001), Avelino et al.\ (2002), Chen 
\& Ratra (2003), Sereno (2003), Bentivegna et al.\ (2004), White et al. 
(2007), Richards et al.\ (2009), and Shafieloo et al. (2010).} Here we 
restate the basic idea from G01 and emphasize some key points that are 
relevant to our analysis.

The basic idea of median statistics is that the true value of a physical
quantity is the median of the set of (error-affected) measurements.
This is based on the assumption that the data set meets two statistical 
requirements: (1) all the measurements are independent; and, (2) there is 
no (overall) systematic error for the whole data set as a group.
In other words, as the number of independent 
measurements goes to infinity, the median will converge to the true value.
The median does not depend on the measurement errors (G01).

Consider a data set consisting of $N$ measurements for a quantity 
that meets the two requirements above. Sort the $N$ measurements from the 
lowest value to the highest and label them $M_i$ respectively, where 
$i = 1,...,N$. Then the probability that the true value for the quantity
 lies between $M_i$ and $M_{i+1}$ is   
\begin{equation}
  P_i = {2^{-N} N!\over i!(N - i)!} ,
\end{equation}
where we set $M_0 = -\infty$  and $M_{N+1}= +\infty$ (G01). The 
range from $M_j$ to $M_{N+1-j}$ (where $j\le~N/2$) defines a confidence 
limit (hereafter c.l.) of $C_j$ percent where 
\begin{equation}
 C_j = 100\times(P_j+P_{j+1}+...+P_{N-j}).
\end{equation} 
The $C_j$'s are a finite number of discrete values, with the number 
depending on $N$. So for any confidence limit commonly used, for example, 
the 95\% c.l., we take the c.l.\ corresponding to the $C_j$ which is the 
smallest among those larger than 95 (G01). These confidence limits do not
depend on the measurement errors (G01). 

Note that the systematic error in the second requirement above is 
different from the individual systematic error quoted as part of the error 
for an individual measurement. If the systematic errors for the individual 
measurements are not correlated, we can treat them as random errors when 
combining individual measurements of a whole data set, as discussed in G01
and below, and the total error can be estimated by studying the histogram 
of the whole data set, without going in to details of the error 
analysis. But if all measurements are affected by the same systematic
shift, i.e., there is a systematic error at the whole 
data set level, a median statistics analysis will give an incorrect result.
This is unlikely to be an issue for the $H_0$ data (G01).
The intermediate case is that a subgroup of data share a similar individual systematic error.
Here we use `subgroup systematic' error 
to denote the part of individual systematic errors that are common to 
all measurements within the subgroup.

Subgroup systematic error is likely the main reason that the first requirement above 
(statistical independence) is not satisfied. One estimate of the 
error contribution from this effect may be derived by dividing the $N$ 
measurements into subgroups that belong to different
measurement techniques (measurements in each such subgroup could very 
likely be affected by similar systematic effects), and then studying the 
differences between results from each subgroup (G01).

\section{Application to Huchra's $H_0$ list}

\subsection{Huchra's $H_0$ list}

The final version of Huchra's Hubble constant measurements list, updated on 
October 7, 2010, contains 553 published estimates (rounded to the 
nearest 1 $\kmsmpc$), some as recent as September 2010. All but three of 
them come with error bars. Most include both statistical and systematical 
errors, although a few have only statistical errors. In this paper we 
use only the quoted central $H_0$ value, and not the error, for each 
measurement. For simplicity, we also use a dimensionless number $h$ instead of 
$H_0$, where 
$H_0 = 100 h\kmsmpc$. 

Aside from some that restate previous results, 
most values on Huchra's list are measurements that are
either from different observations (different raw data), 
different data processings (including calibration and correction), 
or different methods (different relation between distance and observable),
and may include different biases. Each of these has an associated 
error\footnote{For the properties of some sources of error see,
for example, G01 and Rowan-Robinson (2009), and references therein.},
any of which may make the final $H_0$ value differ significantly.
There are examples where the same observations and the same estimation 
technique results in differences as large as two standard deviations 
(see, for example, Rowan-Robinson 2009). The complexity of error sources, 
and the difference in systematic errors estimated by different workers 
in the field, make it a worthy goal to use median statistics to derive a 
summary estimate of $H_0$ from Huchra's list.\footnote{Other techniques
for analyzing heterogeneous collections of measurements, with possibly 
different systematic errors, can be found in, e.g., Press (1997), Bayesian 
method; Podariu et al.\ (2001) and Tammann et al.\ (2008), error 
weighted averaging; and, Freedman \& Madore (2010), both Bayesian and 
frequentist methods.}

All but one measurement in Huchra's list have a primary type label that
indicates method used, and less than half of them also have a secondary 
type label that indicates the ``research group'' involved. For four
measurements in the list the type labels in Huchra's list file look 
ambiguous. We
picked their type labels according to our understanding of both the 
corresponding references as well as Huchra's definition of types. 
(The revised list file is available upon request.)
The primary type provide a simple, but quite likely typical (G01), 
criteria for the subgroup study. For 
conciseness, we focus on the primary type classification in the text, 
mentioning secondary type results (shown along with the primary type 
results in Table 1 and Fig.\ 1) only when necessary.
More sophisticated classification schemes require a careful analysis of
the systematic effects, which is beyond the purview of this paper.

\subsection{Analysis of the complete list}

A median statistics analysis of the 553 $H_0$ values results in a median 
$h=0.68$ and 95\% statistical confidence limits of $0.67 < h < 0.69$. 
Ideally, $h = 0.68 \pm 0.01$ can be quoted as the expected value and 
corresponding 2$\sigma$ error. However, caution is in order 
when quoting these because the two requirements of median statistics are 
very likely not fully met by this $H_0$ list. There are two main 
concerns here: systematic errors that are shared by some of the 
measurements (those in a subgroup) and the restating of prior results 
(in proceedings and summary papers) which result in `restating' 
correlations. Both of these effects make measurements in the list 
statistically dependent. Since we do not make use of the error 
information from the list, we will refer to these effects as (subgroup) 
systematic errors. As mentioned in Sec.\ 2, we can only check the 
reliability of the above results by studying the effects of subgroup
systematics, since we choose to ignore the individual errors.

We first group the 553 measurements into 18 subgroups according to 
the primary type label value in Huchra's list file.
The size of each subgroup and the corresponding median statistics results
are shown in Table 1, while Fig.\ 1 shows the histograms for all but the 
two smallest primary type subgroups. Notice that the subgroup medians 
are different, and many differences between two subgroups are larger 
than half of the 95\% confidence range 
of either group. It is fair to say that most 
subgroups have a `subgroup systematic' error which is close to their median 
minus the true value,
and, within each subgroup, statistical errors result in different 
measurements having different values. Clearly, systematic errors for 
different subgroups have different signs and different values. Following
G01, from the line labelled `Subgroup medians' in the Table, we see that 
$\pm 5.5$ $\kmsmpc$ is a reasonable estimate of the $\pm 95$\% systematic
errors. Furthermore, considering the debates about systematic errors in 
this field, it is possible that the subgroup systematic errors are complex 
enough that we can consider them as pseudo-random errors at the level of 
the whole list (G01). In this case we can use these 18 subgroup medians to 
estimate the overall uncertainty.\footnote{This is qualitatively different 
from the procedure of G01 where the subgroup medians are used to estimate 
the overall systematic uncertainty. Here we think that the median 
statistics of the subgroup medians will give the overall uncertainty 
including both the systematic uncertainty and the statistical uncertainty.
However, since the statistical error here is significantly smaller than 
that determined in G01 (from a smaller set of measurements) while the 
systematic error has not changed significantly, resulting in the systematic
error becoming even more dominant, the procedure adopted by us does not 
result in a quantitatively different total error bar compared to what 
the G01 prescription would give.}
The result is $h = 0.68 \pm 0.055$
(95\% total error).\footnote{As an alternate estimate, for 
those concerned about the reliability of early measurements, we can
estimate a systematic uncertainty by using the latest measurement of 
each primary type. These are listed in the Table 1 column labelled 
`Newest', and the corresponding results are listed in the row labelled 
`Newest values', with a 2$\sigma$ total error of $\pm 6.5$ $\kmsmpc$.}
This may be quoted as a conservative constraint on the Hubble constant 
since we are pretty sure that the 18 measurements (i.e., the subgroup 
medians) are statistically independent. 

Now the `All data' result has an extremely small uncertainty (95\% 
confidence level) range of 2 $\kmsmpc$, while the `Subgroup medians' 
has a relatively large one, 11 $\kmsmpc$. But since the number of 
measurements ($N$) affects the 
uncertainty estimate in a manner similar to the $1/\sqrt{N}$ factor 
in mean statistics (G01), we can not simply conclude that the larger 
error estimate includes more uncertainty information (the so-called 
systematic errors, G01) than the smaller one. To examine the 
effect of subgroup size on the uncertainty estimate we perform a 
simulation. The simulation randomly regroups the 553 measurements 
into 18 subgroups of the same sizes listed in Table 1, and computes the 
median statistics of the subgroup medians of these new subgroups. Regrouping 
100 times results in 100 sets of median statistics values (each set 
consists of the median of 18 subgroup medians and corresponding 95\% 
confidence limits). The median of the 100 95\% c.l.\ ranges is 5 $\kmsmpc$, 
with the largest one equal to 10 $\kmsmpc$. So we see that the uncertainty 
estimates from the 18 group medians in Table 1, 11 $\kmsmpc$, does 
indeed include a source of systematic error. This confirms that it is 
reasonable to use $\pm 0.055$ as the 95\% total error.

\subsection{Analysis of subsamples of the list}

One concern regarding the above results is that the median of `All data', 
$h = 0.68$, may be effected by subgroup systematics. To check this, we 
perform a median statistics analysis of truncated lists of measurements, 
truncated by excluding one subgroup of measurements at a time. We only 
exclude the largest subgroups, since excluding a subgroup with only a few 
measurements does not result in a discernible change. The results are 
shown in the right hand part of Table 1. We see that excluding any single
subgroup does not significantly alter the median and c.l.\ range, at least
in comparison to the 95\% total error above. We ignore the `No 2nd type' 
results here because it is not really a type and also has too many 
measurements included. The only suspicious cases are that excluding the 
`Global Summary' set, where the 95\% c.l.\ range expands most, from 
2 to 4 $\kmsmpc$, and that excluding the `Sandage' set, where the median 
changes most, from 68 to 70 $\kmsmpc$. 

The two subgroups picked out above have relatively smaller 95\% c.l.\ 
ranges. Another subgroup that has a similar small 95\% c.l.\ range is the 
`Key Project' 
type. We also point out here that the `Global Summary' type includes 
results from many summary papers, and is likely the main contributor to 
`restating' correlations. If we look at the histograms of the subgroups 
(Fig.\ 1), we see that, except for these three subgroups, the 
scatter within each subgroup is pretty large compared to 2 $\kmsmpc$, 
the 95\% range for the 553 measurements, even after considering the 
approximate $1/\sqrt{N}$ factor effect. That may explain why none of 
these subgroups affect the median of `All data' significantly. As a further 
check, we construct a subsample 
that contains all the measurements except those belonging to either the 
primary type `Global Summary' or the secondary type `Key project' or 
`Sandage'. There are 362 measurements in this subsample and the median and  
95\% confidence limits are 68 and $66\sim69$ $\kmsmpc$. 

Another consistency check is a ``historical'' 
analysis (G01). Here we consider two subsamples. One, `HST era' set, 
only includes the 367 measurements post 1996, the other, `post G01' set, 
only includes the 196 measurements added to Huchra's list after G01.
The corresponding results are 67 and $65\sim69$ $\kmsmpc$, and 69 
and $67\sim70$ $\kmsmpc$, respectively. Note that there are 13  
papers in Huchra's list added after G01, although they predate G01.
These are not included in the `post G01'.
As a reference, we also compute for two more subsamples, `pre-HST' 
and `pre-G01', that are the complements of the above subsamples 
respectively. The pre-HST set gives 71 and $67\sim75$ $\kmsmpc$, while the 
pre-G01 set gives 67 and $65\sim69$ $\kmsmpc$, identical to the G01
result. 

As a consequence of these consistency checks, we believe that the `All data' 
median and the 95\% c.l.\ range of `subgroup medians' are fairly robust and 
together provide a reasonable summary estimate of $H_0$.

\section{Conclusion}

We use median statistics to study Huchra's list of 553 Hubble constant 
measurements. Ignoring the errors associated with individual measurements,
and assuming there is no systematic error at the whole list level, we 
determine a constraint on $H_0$. We use the median of the complete list 
and estimate the error by only sampling one value, the median, from every 
primary type of subgroup, in an attempt to eliminate any possible 
correlations. This constraint is $H_0 = 68 \pm 5.5$ $\kmsmpc$, where the 
95\% error bar includes both systematic and statistical errors. By 
studying various data subsets, we argue that this result is robust 
and so should be used as a summary estimate of $H_0$.

However, without diving into the detailed systematics of each 
measurement, the statistical independence required by the median statistics
technique can not be conclusively established for the `All data' set. 
Nevertheless, given the complexity of the systematic errors associated 
with measuring distances (as evidenced by the heated debates about them), 
we believe that the above constraint is a reasonable summary value. It is 
probably significant that this lies in the middle of the `low'
Tammann et al.\ (2008) value of $H_0 = 62.3 \pm 1.3$ $\kmsmpc$
and the `high' Freedman \& Madore (2010) value of $H_0 = 73 \pm 4.5$ 
$\kmsmpc$ (both 1$\sigma$ errors).

We thank J.\ Huchra for his list of $H_0$ measurements. We acknowledge 
helpful discussions with R.\ Gott and M.\ Vogeley and financial support 
from DOE grant DE-FG03-99EP41093, NSFC grant 10903006, and
National Basic Research Program of China (973 Program) grant No.2009CB24901.

\clearpage
\begin{deluxetable}{lrlrlrrl}
\tabletypesize{\small}
\tablecolumns{8}
\tablecaption{Hubble Constant Medians (in $\kmsmpc$) by Type}
\tablewidth{0pt}
\tablehead{
\colhead{} & \multicolumn{4}{c}{subgroup of the type}&\multicolumn{3}{c}{subgroup excluding the type}\\
\cline{2-5} \cline{6-8} \\
\colhead{Type of Estimate} &
\colhead{Number} &
\colhead{Median} &
\colhead{95\% c.l.(range)\tablenotemark{a}} &
\colhead{Newest} &
\colhead{Number} &
\colhead{Median} &
\colhead{95\% c.l.(range)\tablenotemark{a}}
          }
\startdata
All data         & 553& 68  &67$\sim$69 \, (2)  &   &&&\\
\tableline
Global Summary   & 111& 70  &68$\sim$72 \, (4)  &73 & 442 & 67 &65$\sim$69 (4)\\
SNe I            & 92 & 64  &60$\sim$65 \, (5)  &64 & 461 & 69 &68$\sim$70 (2)\\
Other            & 83 & 68  &60$\sim$71  (11)  &72 & 470 & 68 &67$\sim$69 (2)\\
Grav.\ Lensing   & 75 & 64  &62$\sim$68 \, (6)  &62 & 478 & 69 &67$\sim$70 (3)\\
Sunyaev-Zeldovich& 46 & 60.5&57$\sim$66 \, (9)  &74 & 507 & 69 &67$\sim$70 (3)\\
B Tully-Fisher   & 23 & 60  &56$\sim$72  (16)  &71 & 530 & 68 &67$\sim$70 (3)\\
IR Tully-Fisher  & 19 & 82  &65$\sim$90  (25)  &60 & 534 & 68 &66$\sim$69 (3)\\
SB Fluctuations  & 18 & 75  &71$\sim$82  (11)  &63 & 535 & 68 &66$\sim$69 (3)\\
Tully-Fisher     & 18 & 72.5&68$\sim$74 \, (6)  &61 & 535 & 68 &66$\sim$69 (3)\\
CMB fit          & 16 & 69.5&58$\sim$72  (14)  &71 & 537 & 68 &67$\sim$69 (2)\\
Glob.\ Cluster LF& 14 & 76.5&65$\sim$82  (17)  &69 & 539 & 68 &66$\sim$69 (3)\\
$D_n-\sigma$     & 10 & 75  &                  &78 &&&\\
I, R Tully-Fisher& 9  & 74  &                  &77 &&&\\
SNe II           & 8  & 59.5&                  &76 &&&\\
Plan.\ Nebulae LF& 6  & 85  &                  &77 &&&\\
Novae            & 3  & 69  &                  &56 &&&\\
Red Giants       & 1  & 74  &                  &74 &&&\\
No 1st Type      & 1  & 85  &                  &85 &&&\\
\tableline
Subgroup medians &    & 71  &64$\sim$75  (11)  &   &&&\\
Newest values    &    &     &63$\sim$76  (13)  &71.5 &&&\\
\tableline
No 2nd type      & 315&69   &67$\sim$70 \, (3)  &76 & 238&67  &63$\sim$70 (7) \\
Cosm.\ depend.\  & 75 &67   &63$\sim$70 \, (7)  &62 & 478&68  &67$\sim$70 (3) \\ 
Sandage          & 71 &55   &55$\sim$57 \, (2)  &63 & 482&70  &69$\sim$71 (2) \\
Key Project      & 62 &72.5 &71$\sim$74 \, (3)  &73 & 491&67  &65$\sim$68 (3) \\
deVaucouleurs    & 21 &95   &80$\sim$99  (19)  &80 & 532&68  &66$\sim$69 (3) \\
Irvine conf.\    & 5  &65   &                  &63 &&&\\
Theory           & 4  &52.5 &                  &72 &&&\\
\enddata
\tablenotetext{a}{We only show the c.l.\ for subgroups with more than 10 
measurements because the c.l.\ for a smaller subgroup is not statistically
reliable. The range is defined as the 
difference between the upper and lower limits.}

\end{deluxetable}

\begin{figure}
\centerline{\epsfig{file=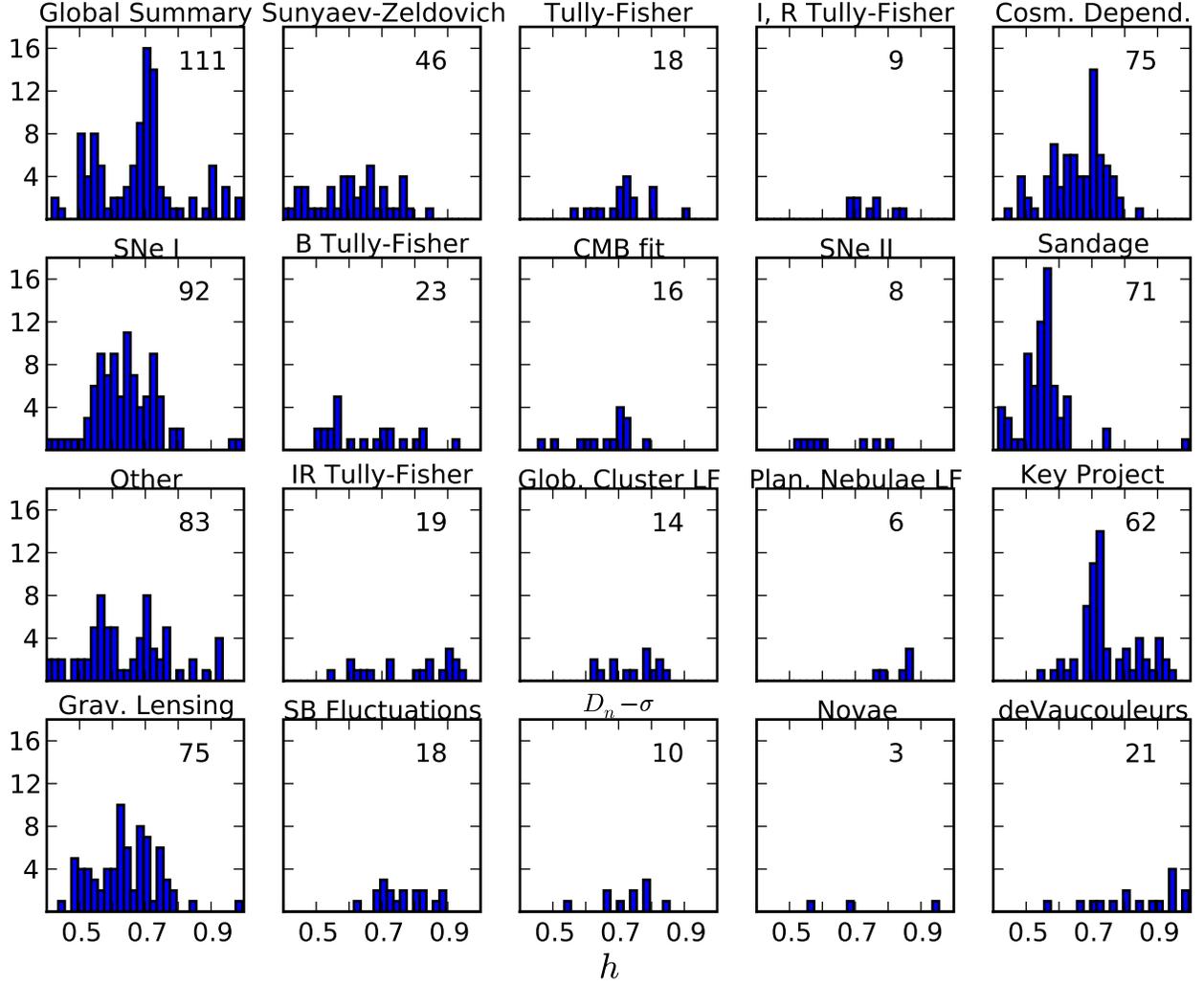,width=20.5cm,angle=0}}
\caption{Distributions of Hubble constant $h$ measurements of the 16 largest 
(of 18) primary and (last column) 4 largest (of 7) secondary type subgroups. 
Each heading lists the primary or secondary type and the number of 
measurements in the subgroup is shown in the upper right hand corner
of each panel. Only measurements in the (horizontal axes) range 
$0.4 \leq h \leq 1$ are shown, but this restriction does not change the 
impression about the distribution of each subgroup.}
\label{f1}
\end{figure} 




\end{document}